\documentclass[aps,preprint,superscriptaddress,groupedaddress]{revtex4}  
\usepackage{dcolumn}   
\usepackage{bm}        
\usepackage{amssymb}   
\usepackage{amsmath}
\usepackage{graphicx,tensor, cleveref}  
\usepackage{dsfont}
\usepackage{cancel}
\usepackage{xcolor}

\begin{document}



\title{2020 Nobel Prize for Physics: \\
Black holes and the Milky Way's darkest secret}
 \author{Joseph Samuel}
 \affiliation{Raman Research Institute, Bangalore-560080, India}
 \affiliation{International Center for theoretical Sciences,\\ Tata Institute of Fundamental Research, Bangalore-560089, India}



\date{\today}

\begin{abstract}
This article was written at the invitation of Current
Science to explain the history and Science behind this year's Nobel prize in Physics.
The article is aimed at a general audience and provides a popular
account and perspective on the subject  of black holes.
\end{abstract}

\maketitle

\section{Introduction}
The Nobel Prize for physics this year was awarded jointly to Roger Penrose,
Reinhard Genzel and Andrea Ghez. The prize recognises theoretical and
experimental advances made in the physics and astronomy of black holes.
Black holes have long fascinated the lay public as well as scientists.
They appear in our language, movies, sitcoms and
science fiction. Scientists are obsessed with them as
they lie at the edge of our current understanding of the Universe.
The scientists honoured by the prize have been instrumental in teasing out the
dark secrets of these mysterious objects. There was a time that the
very existence of black holes was seriously doubted. With advances
in science, technology and understanding,
we now realise that black holes are commonplace
in the Universe. Every galaxy worth the name has one. Indeed, there's one right in our backyard,
in the Milky Way. While their existence is firmly established, black holes continue to throw
up challenges to our understanding of the Universe. The physics and astronomy of black holes
remains a vibrant field. The prize recognises the advances which have
already been made in this field and provides a stimulus for those which
are yet to come.

Roger Penrose (b. 1931)
is a theoretical and mathematical physicist at the University of
Oxford, UK. He has done seminal work in Einstein's General Theory of Relativity. He has also thought
deeply about the foundations of quantum mechanics and its relation to relativity. His interests include the philosophy of science and its popularisation.
He also enjoys recreational mathematics and shares his enthusiasm with the lay public.  As his drawings on the blackboard and
illustrations in his writings reveal, he is also no mean artist.

Reinhard Genzel (b. 1952) is an astrophysicist at the Max Planck Institute
for Extraterrestrial Physics, Garching near Munich.
He also has affiliations with the Ludwig Maximillian University in Munich
and the University of California, Berkeley, USA.

Andrea Ghez (b. 1965) is an astronomer at the
University of California, Los Angeles, USA. She has devoted her life to
understanding the center of our Galaxy, the milky way. She is the fourth
woman to win the Nobel Prize in Physics.

Half of the Nobel Prize is awarded to Roger Penrose and the other half
shared between Reinhard Genzel and Andrea Ghez.

In the rest of this article we set this year's  prize in historical perspective, describe the genesis of the idea of
a black hole, the sustained scepticism that delayed wide acceptance of the idea,
accumulation of experimental evidence for their reality and some of
the challenges that remain for fundamental physics and astronomy.

\section{Prehistory of black holes}
The idea of a black hole occurred independently to
two polymaths- John Mitchell and Pierre-Simon Laplace- separated in
time by thirteen years and in space by the English channel.
John Mitchell was an English clergyman well versed in astronomy, geology,
optics and gravitation.
He was working within the Newtonian scientific paradigms
of his day: the Newtonian theory of gravity, corpuscular light and a geometric interpretation of mechanics.
Mitchell (1783)  based his reasoning \cite{johnmitchell} on the idea of escape speed, the speed
needed for a projectile to escape to infinity from
the surface of an astronomical object. This speed is about 11 km/sec
for the earth and about  600 km/sec for the Sun. These speeds are far less than the speed of light which travels at 300,000 km/sec.
However, if there is a body with same density as the Sun and about 500 times its diameter, even light
would not be able to escape. Thus, he concluded that the most massive bodies
in the Universe may be invisible! We may only be able to deduce their existence by their gravitational influence on other stars. To quote
John Mitchell,
``.. we could have no information from light; If any other luminous bodies would happen to
revolve around them we might still perhaps from the motions of these revolving bodies infer
the existence of the central ones with some degree of probability.'' As we will see below, this is precisely what has been done
by the astronomers Genzel and Ghez of this year's prize.

The same idea was independently conceived \cite{laplace} in France by Pierre-Simon de Laplace in 1796.
Laplace, sometimes called the French Newton, worked on topics as diverse as astronomy, probability, surface tension and
the origin of the solar system. In his ``The System of the world'' he stated: ``Therefore there exists, in the immensity of space, opaque bodies
as considerable in magnitude and perhaps as equally numerous as the stars''. Laplace's arguments were mathematically more sophisticated than his contemporaries,
using differential calculus in contrast to the geometric arguments of John Mitchell. But the content was the same: the Universe may contain ``dark stars''.
Or in Laplace's words, ``The largest bodies in the Universe may well be invisible by reason of their magnitude''.

These prescient speculations lay in obscurity for over a century and  were only revived in the 20th Century with Einstein's General Theory of
Relativity.

\section{Black holes in General Relativity: an idea delayed}
In 1915, Einstein proposed his General Theory of Relativity (GTR) which explained gravitation as the curvature of spacetime.
In about a month, Karl Schwarzschild discovered \cite{schwarzschild} the most
elementary solution of Einstein's equations, which described the spherically symmetric gravitational field of a point mass $M$. The solution had
the feature that the curvature grew without bound as one approached the centre of symmetry, a feature that mathematicians describe as a ``singularity''.
This was not entirely unexpected. The Newtonian gravitational
field of a point mass also grows without bound as
one approaches the centre of symmetry.
However, there was also something intriguing happening at a
finite radius $r=\frac{2GM}{c^2}$,
where $G$ is Newton's gravitational constant and $c$ the speed of light.
This was not properly understood at the time. Some of the expressions describing the geometry seemed to vanish and others to blow up, evading a clear physical
interpretation. There seemed to be some confusion between space and time at this ``Schwarzschild radius''.
Researchers named it the Schwarzschild ``Singularity'' and moved on. They could afford to do so. In any known astronomical body, the problem did not appear,
as the Schwarzschild singularity was deep inside the body, where the vacuum Schwarzschild solution did not apply. For instance, the Schwarzschild radius of the Earth
is 1 cm and that of the Sun is 3 km. The problem would only appear if the body collapses gravitationally to within its Schwarzschild radius.
This means squeezing the earth to within the size of a marble. Or the Sun to within the size of a small town.

In due course, the problem did appear. Arthur Eddington was a renowned astronomer at Cambridge who worked on the life
and death of stars. It was known then that stars shine by burning nuclear fuel in their cores. The heat released in these reactions causes molecular motions and
generates pressure which supports the star against its own gravity. Over millions of years the star shines and exhausts its nuclear fuel. It then cools
and in the absence of thermal molecular motions, contracts under its own gravity. A star with the mass of the Sun would contract till it becomes
a White dwarf, a star not much bigger than the Earth,
which is supported by the quantum mechanical motions of electrons required by the uncertainty principle. Eddington and most of
his contemporaries believed that spent stars would find their final repose as white dwarfs.

This peaceful state of affairs was disturbed by a young Indian, a graduate of Presidency college, Madras who was then working in Cambridge:
Subrahmanyam Chandrasekhar. Chandrasekhar applied himself to understand the equilibrium of White dwarfs. In 1931, he discovered that as the mass of the
star gets bigger, the electrons have to move faster and faster to exert enough pressure to counteract gravity. However, the motion of electrons is
limited by the speed of light and if the mass of the star exceeds 1.4 times the solar mass, gravity wins out and the white dwarf is unstable to gravitational
collapse. Eddington refused to believe this conclusion. He did not have a scientific argument against Chandrasekhar's reasoning, but only a conviction that
``Nature could not behave in this absurd fashion''! Nevertheless, Eddington's eminence and authority prevailed and
the idea of a black hole lay on the shelf for a few more years.

Was there any force that could stop this gravitational collapse? The answer came in
the late nineteen thirties. A paper in 1938 in Zeitschrift \cite{datt} f\"ur Physik by B. Datt, Presidency college Calcutta, gave general solutions of Einstein's equations
in spherical symmetry. Datt was more interested in the cosmological context, but the paper was noticed by the Russian school around L.D. Landau.
Viewed in time reverse, these cosmological solutions can be interpreted as the interior view of gravitational collapse.
Oppenheimer and Snyder (1939) at the University of California, Berkeley
found solutions \cite{oppenheimer} of Einstein's equations that showed that a collapsing star would
continue to collapse past its Schwarzschild radius.  Oppenheimer and Snyder clearly interpret their equations:
``The star thus tends to close itself off from any communication with a distant observer; only its gravitational field persists. ''
They had correctly identified the event horizon of a black hole.
The conclusion was right but the timing was poor. Shortly afterwards, World War II broke out.  Scientific research projects were shelved
in favour of the urgent demands of the war effort.  Many scientists were involved in the war
effort, some of them developing radar to detect enemy planes.
Oppenheimer dropped his research on gravitational collapse and went on to lead
the Manhattan project. Another delay!

Even in 1939, Einstein was unconvinced by the physical reality of the Schwarzschild radius.  Writing in the Annals of Mathematics in 1939,
he argued that time would stand still at the Schwarzschild radius, a patently absurd conclusion. Also bodies falling into the Schwarzschild
radius would appear to hover at the Schwarzschild radius, frozen in time for ever. He seriously doubted that these predictions of
his theory had any physical validity.  He believed
that these absurdities were artefacts of the idealisation involved in a point mass. After the famous point mass solution,
Schwarzschild had in fact discovered a interior spherically symmetric solution of Einstein's equations. But as Schwarzschild had
assumed an incompressible fluid interior, Einstein was able to dismiss this too as an artefact of unreasonable assumptions: in an incompressible
fluid, sound would propagate instantaneously, violating relativity.

The high degree of symmetry of the Oppenheimer-Snyder collapsing solution was also a matter of concern. Perhaps the collapse and resultant
singularity was a  consequence of the artificial initial conditions. Perhaps deviations from spherical symmetry would prevent gravitational collapse.
Maybe the matter would ``bounce back'' due to other interactions, or release its energy  in  a burst of gravitational radiation.
Many cosmological solutions of Einstein's equations were known. Most of these had singularities either in the past or in the future.
Naturally, all the solutions found analytically had a high degree of symmetry, since Einstein's equations are hard to solve without
the simplifying assumption of symmetry. The question remained: are total gravitational collapse and singularities generic in General
Relativity?

\section{Radio astronomy and relativistic astrophysics}
After the war, the world was beating its swords into ploughshares. Radio engineers  turned their war-time instruments
to the skies and discovered a whole new window to the Universe. Unlike light,
radio waves are not absorbed by dust in the Galaxy and radio telescopes  were able to ``see'' clear and deep into the cosmic distance.
Radio waves were detected from the Milky Way and from other galaxies. A new branch of astronomy was born. As the observations mounted, radio astronomers realised that
the skies were not as placid as they appeared. Radio waves were emanating from very tiny regions in the sky. The size of the emitting region had to be less
than a light day since the intensity sometimes changed in a few hours. But the spectral lines of these sources showed that they had to be at an enormous cosmic
distance. From the energy received at the telescopes, it was clear that some of these point like ``stars'' were shining brighter than whole galaxies.
They were called quasars or quasi-stellar objects (QSOs) or Active Galactic Nucleii (AGN). Many sources had radio jets, matter ejected at relativistic
speeds from the centre (Fig.1).
There were violent and energetic events taking place at the centers of galaxies. Tremendous amounts
of energy were being released from an object not much larger than our solar system.

\begin{figure}[h!t]
\includegraphics[scale=0.7]{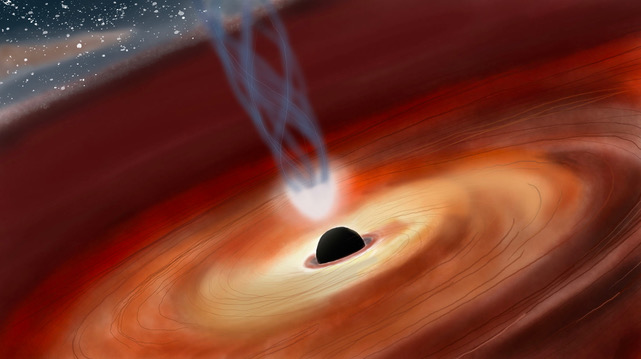}
\caption{Figure shows an artist's impression
of a black hole surrounded by an accretion disc emitting a radio jet.
Figure Credit \copyright Roshni Rebecca Samuel
}
\end{figure}

What was the source of this energy?  Nuclear energy
was an unlikely candidate. Nuclear reactions have about a one percent efficiency in converting mass into energy ($E=mc^2$!). It would need
enormous amounts of matter to be concentrated into a tiny volume in order to explain the observed energy output. Such concentrations of matter
would actually signal a situation close to gravitational collapse. In such compact objects, the gravitational energy of falling matter could
be converted into radiation with an efficiency as high as five percent (for a non rotating black hole). As more data came in, it was clear that many galaxies had
compact objects in their cores. Relativistic astrophysics was born. The Texas symposium in 1963 marked a coming together of mathematics, Physics
and astronomical observations.

\section{Relativity after Einstein}
The period after Einstein's death in 1955 marked a new phase in the development
of general relativity. This was the beginning of mathematical relativity. Much of the earlier confusion with ``Singularities''
stemmed from  the fact that researchers were not sufficiently sophisticated mathematically. They had an undue faith in the coordinates
they used to describe space and time. Some ``singularities'' are only apparent; they are failures of the coordinate system not features
of the spacetime. This is best explained by using geography. Over the Earth, we use latitudes and longitudes to pin point locations.
However, this method fails at the poles of the Earth, as the longitude is ill defined.But there is nothing special about the poles. This is a removable singularity,
a failure of the coordinate system rather than an intrinsic feature of the Earth.
What was needed was an appreciation of global methods, which mathematicians had developed for their own purposes. By patching together local descriptions using coordinates,
they were able to arrive at a global notion: that of a  manifold. With increased sophistication, it becomes clear that the Schwarzschild ``singularity'' is only apparent.
There {\it is} something funny happening at the Schwarzschild radius, but it is not a singularity. It is the event horizon of a black hole!

In the year of Einstein's death in 1955, Amal Kumar Raychaudhuri, working in Calcutta, studied the motion of a cloud of dust in general relativity
and derived an equation (\cite{raychaudhuri}) the Raychaudhuri equation) which was to prove central in further developments.
The equation expresses the attractive nature of gravity: that neighbouring particle trajectories
would tend to focus.
Slightly later, a version of the Raychaudhuri equation was independently derived by L. D. Landau in the Soviet Union.
Two of his colleagues,
E.M Lifshitz and I.M Khalatnikov studied \cite{lifshitz} the question of whether singularities were generic in general relativity. Their conclusion
was negative: singularities were artefacts of symmetry and did not appear generically in the theory.
\color{black}
Roger Penrose was very sceptical of this conclusion. In seminal papers,
he applied the Raychaudhuri equation, using some
global mathematical techniques from differential topology and geometry to arrive \cite{penrose} at the opposite conclusion:
generic solutions of Einstein's equations contain singularities
either in the past (like the big bang) or in the future (as in the centre of a black hole). One of the key ideas he introduced
was the global notion of a trapped surface. Imagine a sphere in ordinary space emitting a flash of light. One part of the wave emanating from the sphere
moves into the sphere contracts everywhere and decreases in area. The other wave  moves out, expands everywhere, increasing in area.
In some gravitational fields, like that inside  the Schwarzschild
radius, it can happen that both these waves (or wave fronts more precisely) are contracting everywhere.
This is what Penrose calls a trapped surface. Once a trapped surface forms, Penrose showed that collapse and singularities are inevitable. Small perturbations
of the trapped surface in the Schwarzschild spacetime do not destroy the trapped surface and it follows that singularities are generic. More precisely, what he showed
was that the spacetime was incomplete. In such a spacetime, photon trajectories would abruptly terminate, suggesting that points of spacetime have been artificially
removed from consideration. Putting them back reveals the singularity. One of the key assumptions in the theorem was that matter was ``reasonable'' {\it i.e} there was no negative mass,
which is the case with all known forms of matter.  Subsequent work by Penrose and Hawking nailed it even more firmly: most spacetimes have singularities, either in the past or in the future.

\section{A blackhole in our backyard}
The Nobel prize was awarded for work by two independent groups, who studied the
central region of our Galaxy over a period of twenty years  and adduced strong evidence for the existence of
a black hole in the centre of our Galaxy.
The estimated mass of the black hole is $4.0$ Million times the mass
of the Sun and corresponds to a Schwarzschild radius of $10^7$ km ($40$ light seconds). The evidence is based on the observations of the motion
of stars in a small region near the center of the Milky way. The observations tell us that the stars are in orbit around a massive
central object. The motion is consistent with Kepler's laws of planetary motion.
The center of the Galaxy is 24,000 light years away from us. At this distance, the motion of stars
appears as a very tiny angular change in the position of the star in the sky. Astronomers refer to this as ``proper motion''.
Detecting the proper motion is a challenging task. First, one
needs to build up a celestial reference frame against which proper motions
can be measured and followed over decades. One also needs fine angular
resolution to distinguish objects in the field of view. In analogy
one needs to be able to distinguish between the left and right eye of a man standing 250 kilometers away.
The angular resolution of a telescope increases with its size and with its operating frequency. (The best angular resolution one can get
with a telescope of size $D$ operating at a wavelength $\lambda$ is
$\frac{\lambda}{D}$ in radians.)
Even with large telescopes operating at small wavelengths, there is a problem because the turbulence
in the earth's atmosphere blurs and distorts the image.

The two groups (one led by Genzel and the other by Ghez), studied the region around the source Sagittarius $A^*$ near the centre of our Galaxy.
This region has fast moving stars and hot ionised gas. The region of interest subtends an angle of only $6^{''}$ (six arc seconds; one arc second is $1/3600$ of a degree).
The theoretical limit on the
angular resolution of a telescope of diameter $10{\rm m}$ operating at the infrared wavelength of $2.2 {\mu {\rm m}}$ is a twentieth of an arc second.  This is called the diffraction limit.
The diffraction limit is hard to achieve because of the turbulent atmosphere, which causes the image to shift on time scales of the order of a second.
To beat this problem, the observers used advanced techniques made possible by modern technology. CCD (Charge coupled devices) enabled larger detector efficiency than the older
photographic plates, permitting shorter exposures. The observations were made with short exposures (about a tenth of a second). Over this timescale, the atmosphere can be regarded
as frozen. The snapshots can then be subject to speckle imaging. In the most basic version, the successive images are combined with a compensatory shift. A more advanced version
called speckle interferometry uses Fourier analysis to produce a high resolution image. Another technique used in imaging is adaptive optics. This method measures the
atmospheric fluctuations in real time and corrects for the distortion of the image. The atmospheric fluctuations have the effect of corrugating the plane
wavefront received from the astronomical source. These corrugations are compensated by having a deformable mirror adapting to cancel the distortion.
Early efforts used a `guide star'': a reference star whose apparent position reveals
the fluctuations of the atmosphere. A more advance version uses a laser to create an artificial guide star in the sky which is used as a reference.

By patient observations \cite{genzel}, \cite{ghez}lasting over decades, the astronomers were able to identify stars orbiting around a central massive object. They were able to
observe the motion of the stars in their orbit and even
complete orbits. Some of the orbits are extremely elliptical and in the closest approach
complete orbits. Some of the orbits are extremely elliptical and in the closest approach
to the central object give us a weak upper bound on its size.
One of the stars has a period of just fifteen years, well within the patience and lifetimes of astronomers.
\color{black}
\begin{figure}[h!t]
\includegraphics[scale=0.8]{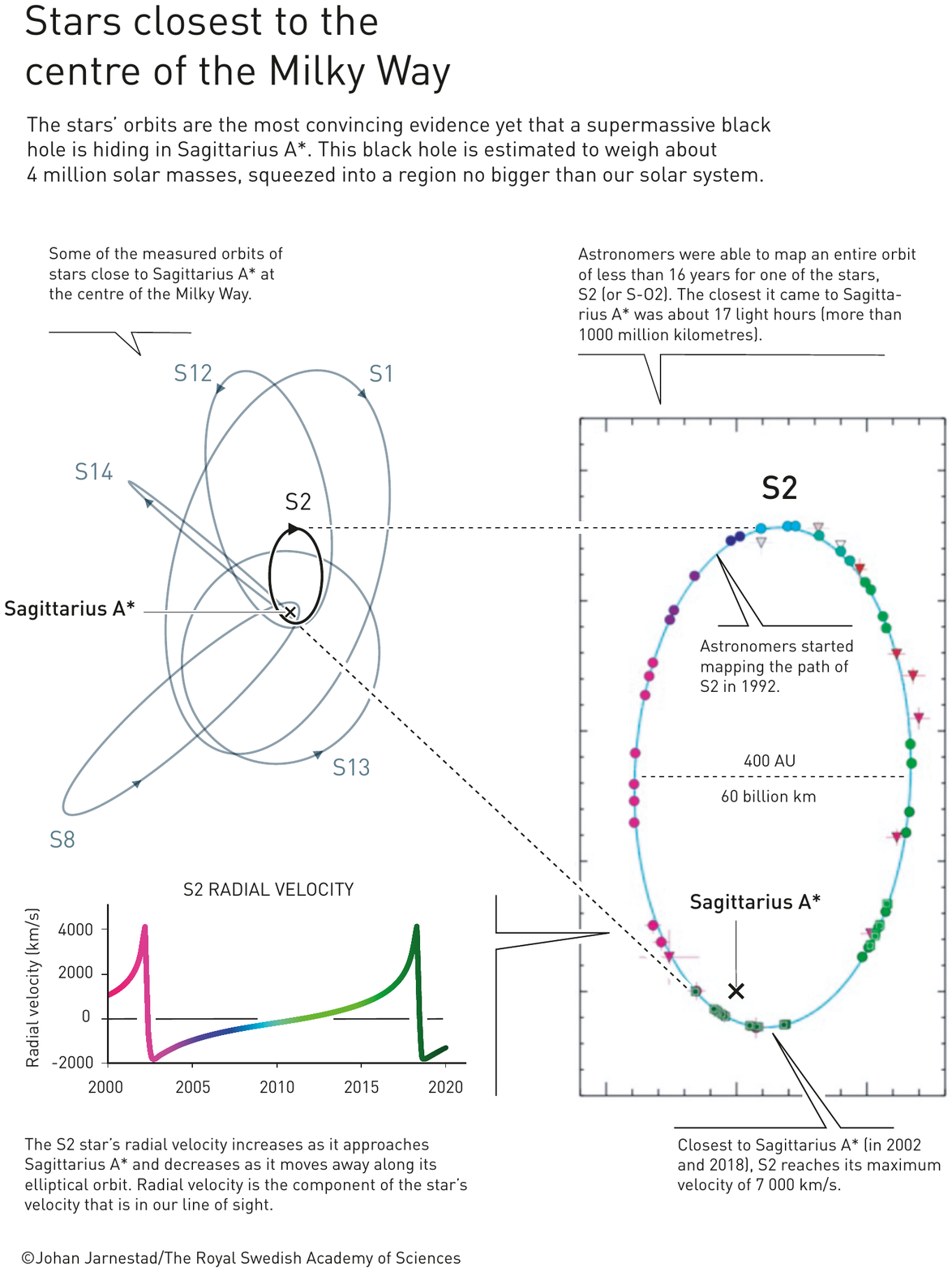}
\vspace{-5cm}
\caption{The figure shows the elliptical orbits of stars around the central object. This figure is the result of decades of observations of stellar motions
in the centre of our Galaxy. Remember that the figure shows the projection of orbits in the plane of the sky.
In projection the focus of the elliptical (or circular) orbit may not lie at the location of the central object.
Figure Courtesy \copyright Johan Jarnestad/The Royal Swedish Academy of Sciences.}
\end{figure}

The stars can
be adequately described by Newtonian gravity. Kepler's laws tell us that in the solar system, planets move in elliptical orbits. What the observers see
(Fig.2) at the centre of our Galaxy is exactly the same, except that it is scaled up in size.
The Sun is replaced by the central black hole, which is 4 Million times the mass of the Sun. The planets are replaced by stars traversing elliptical
orbits. Tracing the orbits of the stars, measuring their positions and velocities tells us the mass of the central object as well as an upper bound
on its size. Nothing fits the bill like a black hole.

\section{Conclusion}
This year's  Nobel prize clearly brings out the interaction between theory and experiment in this area of physics.
The speculations of Mitchell and Laplace  and their illustrious successors were all purely theoretical. It was only with
the coming of Radio astronomy in the 1950's that experimental evidence began to emerge.  The radio observations of Active galactic nucleii
showed that many galaxies have black holes in their centres. The question naturally arose: does the Milky way have one?
Today we not only know the Milky way harbours a black hole, we have measured its mass! The day is not far off when we will
also know its spin!
The experiments now throw up new challenges for the theory of galactic evolution.  Black holes are believed to be
important in the evolution of galaxies. The observations reveal that there are a surprising number of young stars in the vicinity
of the black hole. One would not have expected this, as black holes tend to tidally disrupt large objects like the gas
clouds which form young stars.

The singularity theorems are important developments in Relativity in its modern phase. Apart from telling us clearly that
general relativity predicts black holes, they also tell us that general relativity will fail at some point.
A new theory will be needed to understand the singularity. Understanding black holes has also given us hints
about the new direction. The irreversible nature of gravitational collapse is reminiscent of other irreversible phenomena in physics
like the increase of entropy and the loss of information. Work by Hawking in the 1970s revealed that black holes are thermal objects
in quantum physics. It seems very likely that the missing piece of this puzzle is a quantum theory of gravity. This is the holy
grail of theoretical physics.

\section{Acknowledgements}
I thank Roshni Rebecca Samuel for her artistic drawing  of a black hole in Figure 1 and for the
portraits of the prize winners in Figure 3-5;
and Rajaram Nityananda, Lakshmi Saripalli, Biman Nath, Supurna Sinha, Ravi Subrahmanyan and Sukanya Sinha
for reading through this article and suggesting improvements.

\begin{figure}[h!t]
\includegraphics[scale=0.2]{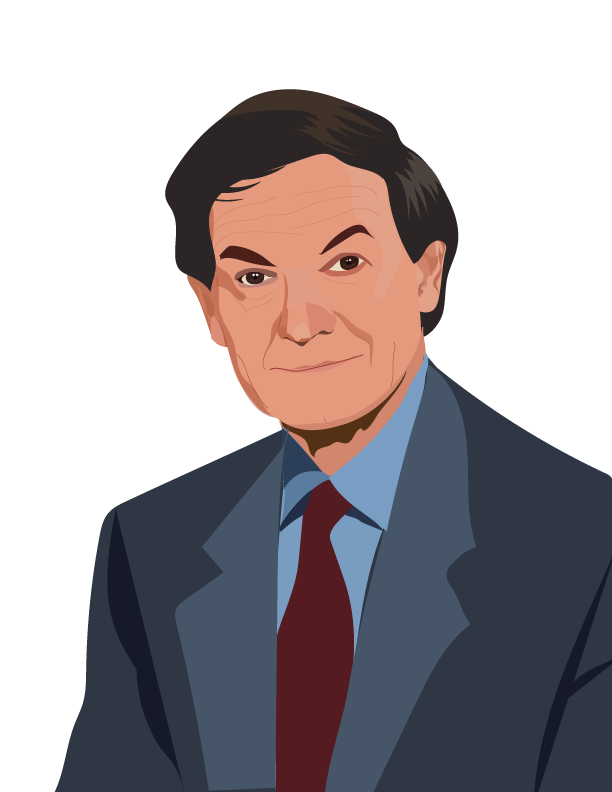}
\caption{Roger Penrose
(Figure Credit \copyright Roshni Rebecca Samuel)
}
\end{figure}

\begin{figure}[h!t]
\includegraphics[scale=0.2]{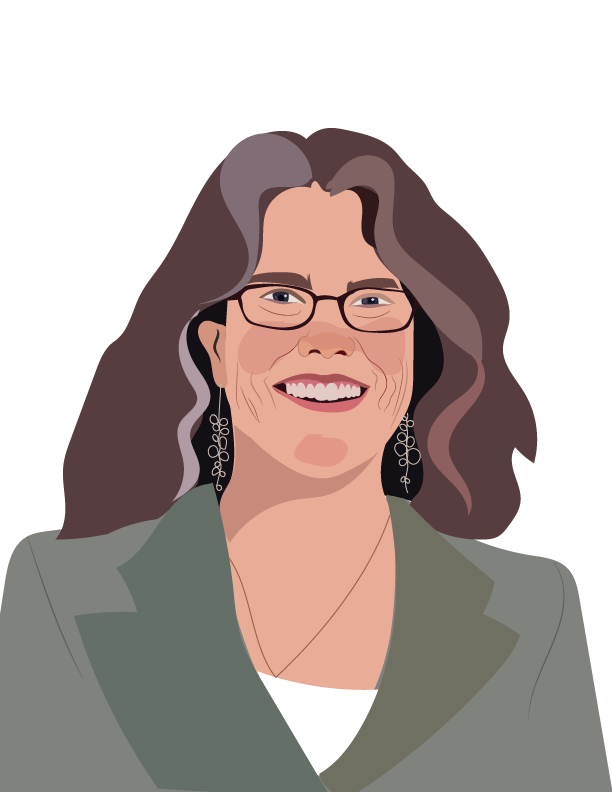}
\caption{Andrea Ghez                                    
(Figure Credit \copyright Roshni Rebecca Samuel)
}
\end{figure}

\begin{figure}[h!t]
\includegraphics[scale=0.2]{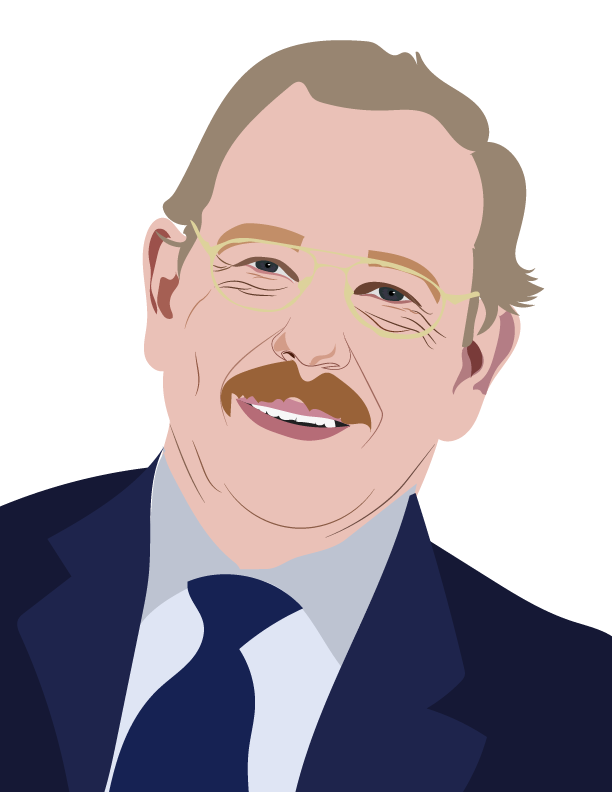}
\caption{Reinhard Genzel   
(Figure Credit \copyright Roshni Rebecca Samuel)
}

\end{figure}

\end{document}